\documentclass[11pt]{article}

\usepackage{amsmath}

\newcommand{\N}{N\raise.7ex\hbox{\underline{$\circ $}}$\;$}

\begin{document}

\title{E.M. Ovsiyuk\footnote{e.ovsiyuk@mail.ru}, V.M. Red'kov\footnote{redkov@dragon.bas-net.by}\\
Coulomb problem for a Dirac particle  in flat Minkowski space and the Heun functions, extension to curved  models
\\[5mm]
{\small  Mozyr State Pedagogical University named after I.P. Shamyakin \\
Institute of Physics, National Academy of Sciences of
Belarus}
}

\date{}

\maketitle

\begin{abstract}

It is shown that there exist several ways to treat the  quantum-mechanical Coulomb problem for a Dirac particle
in flat Minkowski space
with the help of the Heun differential equation, Fuchs's  equation with four singular points.
When extending the problem to curved  spaces of constant curvature, Lobachevsky $H_{3}$  and Rimann $S_{3}$,
  there arise 2-nd order
differential equations of the Fuchs type with 6 singular points, a method to get relevant  equations  with five
singular points has been elaborated.

\end{abstract}

Let us start with a Dirac equation written in spherical tetrad of flat Minkowski space (more detail see in \cite{Book-2})
$$
\left [ \;i \; \gamma ^{0} \; \partial_{t} \;   + \; i \;  ( \gamma ^{3}
\; \partial_{r} \; + \; { \gamma ^{1} \; \sigma ^{31} \; + \;
\gamma ^{2} \; \sigma ^{32} \over  r } ) \;  + {1 \over r} \;
\Sigma_{\theta \phi} \;  - \; m \; \right ]\;  \Psi (x)  =  0    \; ,
$$
$$
\Sigma _{\theta ,\phi } \; = \;  i\; \gamma ^{1} \partial
_{\theta} \;+\; \gamma ^{2} \;  {\;  i \partial _{\phi} \; + \;
 i\; \sigma ^{12} \over \sin \theta } \;  .
$$

\noindent In that basis, spherical  solutions are constructed on the base of
a substitution
(Wigner's $D$ functions are noted as
$D^{j}_{-m,\sigma }(\phi ,\theta ,0)  \equiv  D_{\sigma}$):

$$
\Psi _{\epsilon jm}(x) \;  = \; {e^{-i\epsilon t} \over r} \;
\left | \begin{array}{l}
        f_{1}(r) \; D_{-1/2} \\ f_{2}(r) \; D_{+1/2}  \\
        f_{3}(r) \; D_{-1/2} \\ f_{4}(r) \; D_{+1/2}
\end{array} \right | \; .
\eqno(1)
$$

\noindent With the use of recurrent relations \cite{Varshalovich-Moskalev-Hersonskiy-1975}
$$
\partial_{\theta} \; D_{+1/2} \; = \;  a\; D_{-1/2}  - b \; D_{+3/2}  \; ,
$$
$$
{- m - 1/2 \;  \cos \theta  \over  \sin \theta } \; D_{+1/2}\; =\;
- a \;  D_{-1/2} - b \;  D_{+3/2}  \; ,
$$
$$
\partial_{\theta} \;  D_{-1/2} \; = \;  b \; D_{-3/2} - a \; D_{+1/2}  \; ,
$$
$$
{- m + 1/2 \; \cos \theta \over \sin \theta}  \;  D_{-1/2} \; = \;
 - b \; D_{-3/2}  - a \;  D_{+1/2}  \; ,
$$

\noindent where
$$
a = {j +  1/2 \over 2} , \qquad b  = {1\over 2}
\;\sqrt{(j-1/2)(j+3/2)}\;,
$$

\noindent we derive  (below   $\nu  = j + 1/2$ )

$$
 \Sigma _{\theta ,\phi } \; \Psi _{\epsilon jm}(x) \;  = \; i\; \nu \;
{e^{-i\epsilon t } \over r} \; \left | \begin{array}{r}
        - \; f_{4}(r) \; D_{-1/2}  \\  + \; f_{3}(r) \; D_{+1/2} \\
        + \; f_{2}(r) \; D_{-1/2}  \\  - \; f_{1}(r) \; D_{+1/2}
\end{array} \right |        .
$$

\noindent and after simple calculations  obtain radial equations

$$
\epsilon   f_{3}   -  i  {d \over dr}  f_{3}   - i {\nu \over r}
f_{4}  -  m  f_{1} =   0  \; ,\qquad \epsilon   f_{4}   +  i  {d
\over dr} f_{4}   + i {\nu \over r} f_{3}  -  m  f_{2} =   0    \;
,
$$
$$
\epsilon   f_{1}   +  i  {d \over dr}  f_{1}  + i {\nu \over r}
f_{2}  -  m  f_{3} =   0  \; , \qquad \epsilon   f_{2}   -  i  {d
\over dr} f_{2}   - i {\nu \over r} f_{1}  -  m  f_{4} =   0  \; .
\eqno(1b)
$$

Usual   $P$-reflection operator
\cite{1980-Landau-4} in Cartesian basis
$\hat{\Pi}_{C}  =  i \gamma ^{0} \otimes \hat{P}$

$$
\hat{\Pi}_{C.}  = \left | \begin{array}{cccc}
          0 &  0 &  i &   0  \\
          0 &  0 &  0 &   i  \\
          i &  0 &  0 &   0  \\
          0 &  i &  0 &   0
\end{array} \right |       \; \otimes  \; \hat{P} \; , \qquad
\hat{P} (\theta , \phi ) = (\pi  - \theta, \; \phi+ \pi )
$$

\noindent after translation to spherical basis

$$
\hat{\Pi}_{sph} = S(\theta ,\phi ) \; \hat{\Pi}_{C}\;
S^{-1}(\theta ,\phi ) \; ,
$$
assumes the form

$$
\hat{\Pi}_{sph} \;  = \left | \begin{array}{cccc}
0 &  0 &  0 & -1   \\
0 &  0 & -1 &  0   \\
0 &  -1&  0 &  0   \\
-1&  0 &  0 &  0
\end{array} \right |
\; \otimes  \; \hat{P} \; .
$$

\noindent From eigenvalue equation

$$
\hat{\Pi}_{sph}  \Psi _{jm} =  \Pi \; \Psi _{jm}, \qquad
\hat{P}   \; D^{j}_{-m,\sigma } (\phi ,\theta ,0)  = (-1)^{j} \;
D^{j}_{-m,-\sigma} (\phi ,\theta ,0)
$$

\noindent   it follows

$$
\Pi = \; \delta \;  (-1)^{j+1} , \;\; \delta  = \pm 1 \; ,  \qquad
f_{4} = \; \delta \;  f_{1} , \qquad  f_{3} = \;\delta \; f_{2} \;
;
$$

\noindent that is

$$
\Psi (x)_{\epsilon jm\delta } \; = \; {e^{-i\epsilon t} \over  r }
\; \left | \begin{array}{r}
     f_{1}(r) \; D_{-1/2} \\
     f_{2}(r) \; D_{+1/2} \\
\delta \; f_{2}(r) \; D_{-1/2}   \\
\delta \; f_{1}(r) \; D_{+1/2}
\end{array} \right |  \; .
\eqno(2a)
$$

\noindent Allowing for (2a), we simplify eqs.   (1b):

$$
({d \over dr} \;+\; {\nu \over r}\;) \; f \; + \; ( \epsilon  \;+
\;
 \delta \; m )\; g \; = \;0 \; ,
 $$
 $$
({d \over dr} \; - \;{\nu \over r}\;)\; g  \;- \; ( \epsilon \; -
\;
 \delta\;  m )\; f\; =\; 0     \; ,
\eqno(2b)
$$

\noindent where instead of   $f_{1}$  Ё  $f_{2}$   new functions are used

$$
f \; = \; {f_{1} + f_{2} \over \sqrt{2}} \; , \qquad g \; = \;
{f_{1} - f_{2} \over i \sqrt{2}} \; .
$$

To simplify the wave functions  $ \Psi
_{\epsilon jm} \rightarrow  \Psi _{\epsilon jm\delta } $ and  one can use
so-called Dirac (or Johnson-Lippmann) operator  \cite{1980-Landau-4}   $\hat{K}$ (see also \cite{1939-Pauli}),

$$
\hat{K} \; = \;  - \gamma ^{0} \gamma ^{3} \; \Sigma _{\theta
,\phi } = \gamma ^{0}\gamma ^{3} \;  \left [\; \gamma ^{1} \; (\partial
_{\theta } + 1/2) +
 {\gamma ^{2} \over \sin \theta } \; \partial_{\phi} \; \right  ]  \; .
$$

\noindent Indeed, from  the eigenvalue equation  $\hat{K} \; \Psi
_{\epsilon jm} (x) = K \; \Psi _{\epsilon jm}$ we get
$$
K = - \delta \; (j+1/2) \;\;  ,\;\;  \delta \; = \pm  1 \; ,
$$
$$
f_{4} = \; \delta \; f_{1}\; ,\qquad f_{3} = \; \delta \;  f_{2}
\; ,
$$

\noindent which coincides with the above restrictions.

Transition to a  Coulomb problem  for a Dirac  particle is reached by formal chane
the radial  system (2b)
$$
 ( {d \over d  r }  +  {\nu \over r } \;) \; f \; + \; ( E  + { e \over  r } \;+ \;
  m )\; g \; = \;0 \; ,
 $$
 $$
 ({d \over d r } \; - \; {\nu \over r  } \;)\; g  \;- \; ( E + {e \over r }
\; - \;
   m )\; f\; =\; 0     \; .
\eqno(3)
$$

Let us perform a linear transformation over functions $f(r)$ and $
g(r)$ (its coefficients may depend on the radial variable, let its
determinant obey an identity  $a (r) b (r) - c (r) d(r)  = 1$)

$$
f (r) = a  \;  F (r) + c  \; G (r)\; , \qquad
 g (r) =  d  \; F(r) + b  \; G(r) \; ,
\; \;
$$
$$
F(r) = b  \; f(r)  - c \;  g (r) \; , \qquad
  G(r) = - d  \; f(r) + a  \; g(r) \; .
\eqno(4a)
$$

Let us combine eqs.  (3) as follows: the first equation is
multiplied by $+b$, the second by  $- c$, and add the results;
analogously, the firs equation multiply by $-d$ , the second by
$+a$, and sum  the results. Thus we arrive  at

$$
\left [ \; {d \over d r}  -b'a +c'd  + {\nu \over r} \; (ba +cd) +
(E+{e \over r} + m) \; bd + (E + {e \over r} -m)\;  ca \; \right ]
F
$$
$$
= \left [ \; b'c - b c' - {\nu \over  r} 2bc - (E+{e \over r} +m)
b^{2} - (E + {e \over r} -m) c^{2} \; \right ]  G \; ,
$$

$$
\left [ \; {d \over d r}   + d'c - a'b  - {\nu \over r} \; (dc
+ab) - (E+{e \over r} + m) \; bd - (E + {e \over r} -m)\;  ca \;
\right ]  G
$$
$$
= \left [ \; -d'a +d a' + {\nu \over  r} 2ad  + (E+{e \over r} +m)
d^{2} + (E + {e \over r} -m) a^{2} \; \right ]  F \; .
$$
$$
\eqno(4b)
$$

For simplicity, let the transformation (2) does not depend on
$r$, and let it be  orthogonal:

$$
S = \left | \begin{array}{cc}
a & c \\
d & b
\end{array} \right | =
\left | \begin{array}{cc}
\cos A/2 &  \sin A/2  \\
- \sin A/2 & \cos A/2
\end{array} \right |\,,
$$

\noindent which simplifies eqs. (3)

$$
\left (  {d \over d r}   + {\nu \over r} \;\cos A  -  m  \; \sin A
\right  )
 F = \left (  - {\nu \over  r} \sin A -{e \over r}  - E   - m \cos A
\right )  G \; , \qquad \;\;
$$

$$
\left (  {d \over d r}     - {\nu \over r} \; \cos A  +  m  \;
\sin A \right )  G =  \left (   - {\nu \over  r} \sin A  +{e \over
r} + E  - m \cos A  \right )  F \; . \eqno(5)
$$

There exist four possibilities (in fact, only two of them are  different)

\vspace{5mm} 1)
$$
 - {\nu \over  r} \sin A  +{e \over r}=0\;, \qquad \sin A = {e \over  \nu }\;, \qquad
 \cos A = \sqrt{1 - e^{2} / \nu^{2}}\; ,
 $$
$$
\cos{A\over 2} = \sqrt{{\nu + \sqrt{\nu^{2} - e^{2}} \over 2
\nu}}\; , \qquad \sin{A\over 2} = \sqrt{{\nu - \sqrt{\nu^{2} -
e^{2}} \over 2 \nu}} \; ;
$$

$1'$)
$$
 - {\nu \over  r} \sin A  -{e \over r}=0\;, \qquad
 \sin A = -  {e \over  \nu }\;, \qquad  \cos A = \sqrt{1 - e^{2} / \nu^{2}} \; ,
 $$
$$
\cos{A\over 2} = \sqrt{{\nu -\sqrt{\nu^{2} - e^{2}} \over 2
\nu}}\; , \qquad \sin{A\over 2} = \sqrt{{\nu + \sqrt{\nu^{2} -
e^{2}} \over 2 \nu}}  \; ; \eqno(6a)
$$

\vspace{5mm} 2)
$$
 E  - m \cos A  = 0 \; , \qquad  \cos A = + {E \over m} \;, \qquad   \sin A = \sqrt{1 -E^{2} / m^{2} }  \; ,
$$
$$
\cos {A\over 2} =  \sqrt{{m+E \over 2m}}\; , \qquad \sin {A\over
2} =  \sqrt{{m-E \over 2m}} \; ;
$$

$2'$)
$$
- E  - m \cos A  = 0 \; , \qquad  \cos A = -{E \over m} \;, \qquad
\sin A = \sqrt{1 -E^{2} / m^{2} }  \; ,
$$
$$
\cos {A\over 2} = \sqrt{{m-E \over 2m}}\; , \qquad \sin {A\over 2}
= \sqrt{{m+E \over 2m}} \; . \eqno(6b)
$$

First, consider the case 1). Eqs.  (5) takes the form

$$
\left (  {d \over d r}   + {\nu \over r} \cos A  -  m   \sin A
\right  )
 F =   \left (  -{2e \over r}  - E   - m \cos A  \right )  G \; ,
$$
$$
\left (  {d \over d r}     - {\nu \over r}  \cos A  +  m   \sin A
\right )  G =  (   E  - m \cos A  ) \; F \; . \eqno(7)
$$

\noindent After excluding the function  $F$, we get a second order
equation for $G$

$$
\left (  {d \over d r}   + {\nu \over r} \cos A  -  m   \sin A
\right   ) \left ( {d \over d r}     - {\nu \over r}  \cos A  +  m
\sin A  \right   ) G
$$
$$
=  (   E  - m \cos A  ) \left (   -{2e \over r}  - E   - m \cos A
\right  )  G \; , \eqno(8a)
$$

\noindent or in  more detailed form

$$
\left (\; {d^{2} \over dr^{2}}  +   E^{2} -m^{2} + {\nu  \cos A  -
\nu^{2} \cos^{2} A \over r^{2} } \right.
$$
$$
\left. + {2eE -
   2e m\cos A  +2m\nu \sin A \cos A \over r}     \right )  G = 0 \; .
$$

\noindent Taking in mind identity
 $\sin A = e/ \nu$,  thus equation reduces to

$$
\left (\; {d^{2} \over dr^{2}}  +   E^{2} -m^{2} + {\nu  \cos A  -
\nu^{2} \cos^{2} A \over r^{2} } + {2eE  \over r}     \right ) G =
0 \; . \eqno(8b)
$$

\noindent After changing the  variable,
 $
x=2\,\sqrt{m^{2}-E^{2}}\,r$, it reads

$$
{d^{2}G\over dx^{2}} + \left(-{1\over 4}-{\nu \cos A\, (\nu \cos
A-1)\over x^{2}} + {eE  \over \sqrt{m^{2}-E^{2}}\,x}\right)G=0\,.
$$

\noindent With the use of  a substitution  $ G (x) = x^{a} e^{bx}
\varphi (x)$,  for $\varphi$ we get

$$
x\,{d^{2}\varphi\over dx^{2}}+(2\,a+2\,b\,x)\,{d \varphi\over dx}
$$
$$
+\left[(b^{2}-{1\over 4})\,x+{a^{2}-a-\nu \cos A\ (\nu \cos
A-1)\over x}+2ab+ {eE \over \sqrt{m^{2}-E^2}}\right]\varphi=0\, .
$$

\noindent When
$$
 a =  +\nu\,\cos A  = \sqrt{\nu^{2} - e^{2}} \,, \qquad
   b=-{1 \over 2}
\; ,
$$

\noindent this equation for $\varphi$ becomes simpler

$$
x\,{d^{2}\varphi\over dx^{2}}+(2\,a-\,x)\,{d \varphi\over dx}
-\left[a-{eE  \over \sqrt{m^{2}-E^2}}\right]\varphi=0\, ,
$$

\noindent which is  a confluent hypergeometric equation

$$
x \;  Y '' +( \gamma -x) Y'  - \alpha Y =0 \; , \qquad \alpha=
a-{eE \over \sqrt{m^{2}-E^2}}\, , \qquad \gamma=2a\, .
$$

\noindent To polynomials there correspond the  know restriction $
\alpha = - n, \; n=0,1,2, ...$, which gives  the  known energy
quantization rule

$$
a-{eE \over \sqrt{m^{2}-E^2}} = -n \qquad \Longrightarrow \qquad E
= { m \over  \sqrt{1 +  e^{2} /  (n + \sqrt{\nu^{2} - e^{2}} )^{2}
}} \; . \eqno(8c)
$$

In turn,  from (7) it follows a second order equation for
 $F$

$$
\left (  {d \over d r}     - {\nu \over r} \; \cos A  +  m  \;\sin
A  \right  ) {r \over    2e   + (E   + m \cos A )r } \left (  {d
\over d r}   + {\nu \over r} \;\cos A  -  m  \; \sin A  \right  )
F
$$
$$
 = (   -E  + m \cos A  )  F \; ,
\eqno(9a)
$$

\noindent or

$$
 \left [ (  {d \over d r}     - {\nu \over r} \; \cos A  +  m  \;\sin A   )
(  {d \over d r}   + {\nu \over r} \;\cos A  -  m  \; \sin A   )
\right.
$$
$$
\left. + {d \over dr } \ln \left ( {r \over    2e   + (E   + m
\cos A )r } \right ) (  {d \over d r}   + {\nu \over r} \;\cos A -
m  \; \sin A   )\right ] F
  $$
 $$
 =( {  2e \over r}   + E   + m \cos A )\;  (   -E  + m \cos A  ) \; F \; ,
$$

\noindent and further

$$
\left [ {d^{2} \over dr^{2}}  +  \left ({1 \over r} - {E+ m \cos A
\over 2e +(E+ m \cos A) r} \right )
  {d \over d r}    \right.
$$
$$
\left. +
  {\nu \over r^{2} } \;\cos A  -  {m  \; \sin A \over r}   -
\left ( {1 \over r} -{2e \over r [ 2e +(E+ m \cos A) r]} \right )
( {\nu \over r} \;\cos A  -  m  \; \sin A   ) \right.  $$
  $$
\left. + E^{2} -m^{2}  -
  {\nu \cos A \over r^{2}} +   {e^{2} - \nu^{2}  \over r^{2}}       +
{2eE \over r}     \right ] F = 0 \;.
$$

\noindent Finally, we arrive at

$$
\left [ {d^{2} \over dr^{2}}  + ({1 \over r} - {E+ m \cos A \over
2e +(E+ m \cos A) r} )
  {d \over d r}   \right.
$$
$$
\left. + E^{2} -m^{2}    +{2eE \over r}  +   {e^{2}  - \nu^{2}
\over r^{2}}   -
  { 2em \sin A + \nu \cos A (E +m \cos A)   \over r \;  [ 2e + (E+ m \cos A) r] }
         \right   ] F = 0 \; .
$$

Let us introduce special designation for the  additional singular
point

$$
-{2e \over E + m \cos A } =  R \; ;
$$

\noindent then we obtain

$$
\left [ {d^{2} \over dr^{2}}  + ({1 \over r} - {1 \over  r - R} )
  {d \over d r}   \right.
$$
$$
\left. + E^{2} -m^{2}    +{  2eE \over r}  +   {e^{2}  - \nu^{2}
\over r^{2}}   +
  {  m R \;  \sin A - \nu \cos A    \over r \;  (  r-R)  }
         \right   ] F = 0 \; .
\eqno(9b)
$$

\noindent After changing the  variable $ x  = r /  R$,  it reads

$$
{d^{2}F\over dx^{2}}+\left[{1\over x}-{1\over
x-1}\right]\,{dF\over dx} +\left[ (E^{2}-m^{2})\,R^{2}-
{\nu^{2}-e^{2}\over x^{2}} \right.
$$
$$
\left. + {-\nu \cos A+m  R\, \sin A\over x-1}+ { 2eRE-m R\sin
A+\nu \cos A \over x} \right]F=0\, . \eqno(9c)
$$

Let us  search solutions in the form $
F=x^{a}\,e^{bx}\,\varphi(x)$, the  function $\varphi$ obeys

$$
{d^{2} \varphi \over dx^{2}}+\left[{2a+1\over x}+2b-{1\over
x-1}\right]\,{d \varphi\over dx}
$$
$$
+ \left[b^{2}+(E^{2}-m^{2})\,R^{2}+ {a^{2}-\nu^{2}+e^{2}\over
x^{2}}-{a+b+\nu \cos A-m R\, \sin A\over x-1} \right.
$$
$$
\left. + {a+b+2ab+R(2eE-m \sin A)+\nu \cos A\over
x}\right]\,\varphi=0\,.
$$

\noindent When  $a,b$ taken according to (below we will use
underlines values)
$$
a=\underline{+\sqrt{\nu^{2}-e^{2}}}\,, \;\; -
\sqrt{\nu^{2}-e^{2}}\; ,
$$
$$
b= +\sqrt{m^{2}-E^{2}}\,R\,,  \underline{ -\sqrt{m^{2}-E^{2}}\,R}
\eqno(10a)
$$

\noindent the above equation becomes simpler
$$
{d^{2} \varphi \over dx^{2}}+\left[2b+{2a+1\over x}-{1\over
x-1}\right]\,{d \varphi\over dx}
$$
$$
+\left[{a+b+2ab+ 2eRE- m R\sin A + \nu \cos A\over x}-{a+b+\nu
\cos A-m R\, \sin A\over x-1}\right]\,\varphi=0\, .
$$
$$
\eqno(10b)
$$

\noindent In can be recognized as a confluent Heun equation for
$G(\alpha,\beta,\gamma,\delta,\eta,z)$

$$
G''+\left ( \alpha+{1+\beta\over z}+{1+\gamma\over z-1}\right )G'
$$
$$
+ \left ({1\over 2}\,{\alpha+\alpha \beta-\beta-\beta
\gamma-\gamma-2\eta\over z}+{1\over 2}\, {\alpha+\alpha
\gamma+\beta+\beta \gamma+\gamma+ 2\delta+2\eta\over z-1}\right )
G=0\, \eqno(10c)
$$

\noindent with parameters

$$
\alpha=2b\,, \qquad \beta=2a\,, \qquad \gamma=-2\,,
$$
$$
\delta=2\,eER\,,  \qquad \eta=1+m R\,\sin A-2\,eER-\nu \cos A\,.
\eqno(10d)
$$

Let us use  the known condition for polynomial solutions

$$
\delta=-\left (n+{\beta+\gamma+2\over 2}\right )\,\alpha\,,\qquad
n=0,1,2,..., \eqno(11a)
$$

\noindent it results the energy  quantization rule

$$
a= + \sqrt{\nu^{2}-e^{2}}\,,\qquad b= - \sqrt{m^{2}-E^{2}}\,R\,,
$$
$$
eER=(n+\sqrt{\nu^{2}-e^{2}})\,\sqrt{m^{2}-E^{2}} \; R \; ,
$$
so we arrive at

$$
E = { m \over  \sqrt{1 +  e^{2} / (n + \sqrt{\nu^{2} - e^{2}}
)^{2}       }} \; , \eqno(11b)
$$

\noindent which coincides with the  known formula for energy
levels.

Let us consider the case  2).  Eq. (5) takes the form

 $$
\left (  {d \over d r}   + {\nu \over r} \;\cos A  -  m  \; \sin A
\right  )
 F = \left (  - {\nu  \sin A  + e \over  r}     - 2m \cos A  \right )  G \;
,
$$
$$
\left (  {d \over d r}     - {\nu \over r} \; \cos A  +  m  \;
\sin A \right )  G =      {e-\nu \sin A \over  r}  \; F \; .
\eqno(12)
$$

One  can obtain a second order equation  for $G(r)$:

 $$
\left [ (  {d \over d r}   + {\nu \over r} \;\cos A  -  m  \; \sin
A   ) \;r (  {d \over d r}     - {\nu \over r} \; \cos A  +  m  \;
\sin A   )   \right.
$$
$$
\left. + ( e - \nu \sin A ) (   { e + \nu  \sin A \over  r}  + 2m
\cos A  ) \right ]  G = 0 \; , \eqno(13a)
$$

\noindent from whence it follows

 $$
\left [   {d \over d r}     - {\nu \over r} \; \cos A  +  m \;\sin
A    + r \left (  {d ^{2}\over d r^{2} }   + {\nu \cos A \over
r^{2} } \;  -
 (    {\nu \over r} \;\cos A  -  m  \; \sin A   )^{2}  \right )
   \right.
$$
$$
\left. + ( e - \nu \sin A ) \left (   { e + \nu  \sin A \over  r }
+ 2m \cos A  \right  ) \right ]  G = 0 \; ,
$$

\noindent or

 $$
\left (    {d ^{2}\over d r^{2} }  +  {1 \over r }   {d \over d r}
-   m ^{2}  \; \sin^{2} A          +     { e^{2} - \nu ^{2} \over
r^{2} }  +
 { 2me  \cos A   \over   r}   +  {m  \sin A  \over   r } \right )  G = 0 \; .
$$

\noindent And further,  taking into account identity
 $\cos A = E /m$, we arrive at

  $$
\left (   {d ^{2}\over d r^{2} }         + {1 \over r }   {d \over
d r}       +
  E^{2} - m^{2}         +     { e^{2} - \nu ^{2}   \over  r^{2} }  +
 { 2 eE   \over   r}    +  {\sqrt{m^{2} - E^{2} }  \over   r }   \right )  G = 0 \; .
\eqno(13b)
$$

Making the  change of variables $x=2\,\sqrt{m^{2}-E^{2}}\,r$,
 we get

$$
{d^{2}G\over dx^{2}} +{1\over x}\,{dG\over dx}+ \left(-{1\over
4}-{\nu^{2}-e^{2}\over x^{2}} +{1\over 2}\, {m^{2}-E^{2}+2 E
e\,\sqrt{m^{2}-E^{2}}  \over (m^{2}-E^{2})\,x}\right)G=0\,.
$$

\noindent Let
 $ G (x) = x^{a} e^{bx} \varphi
(x)$,  the function $\varphi$ satisfies
$$
x\,{d^{2}\varphi\over dx^{2}}+(2\,a+1+2\,b\,x)\,{d \varphi\over
dx}
$$
$$
+\left[(b^{2}-{1\over 4})\,x+{a^{2}-\nu^{2}+e^{2}\over
x}+2ab+b+{1\over 2}\, {m^{2}-E^{2}+2 E e\,\sqrt{m^{2}-E^{2}} \over
m^{2}-E^{2}}\right]\varphi=0\, .
$$

\noindent When

$$
 a =   \sqrt{\nu^{2} - e^{2}} \,, \qquad
   b=-{1 \over 2}
\; ,
$$

\noindent we get

$$
x\,{d^{2}\varphi\over dx^{2}}+(2\,a+1-x)\,{d \varphi\over
dx}-\left (a-\, { E e\, \over \sqrt{m^{2}-E^{2}} }\right
)\varphi=0 \; ,
$$

\noindent which is a confluent hypergeometric equation

$$
x \;  Y '' +( \gamma -x) Y'  - \alpha Y =0 \; , \qquad \alpha=
a-\, { E e\, \over \sqrt{m^{2}-E^{2}} }\, , \qquad \gamma=2a+1\, .
$$

\noindent Solutions become polynomials if $ \alpha = - n, \;
n=0,1,2, ...$, this provides us with the energy spectrum

$$
E = { m \over  \sqrt{1 +  e^{2}  /  (n + \sqrt{\nu^{2} - e^{2}}
)^{2}        }} \; . \eqno(13c)
$$

In turn, from (12) it follows a second  order equation for
 $F(r)$

$$
\left (  {d \over d r}     - {\nu \over r} \; \cos A  +  m  \;\sin
A \right ) { r \over  \nu \sin A + e + 2m\cos A \; r } \left (  {d
\over d r} + {\nu \over r} \;\cos A  -  m  \; \sin A  \right  )  F
$$
$$
=     {\nu \sin A  - e \over  r}  \; F \; , \eqno(14a)
$$

\noindent that is

$$
\left [ (  {d \over d r}     - {\nu \over r} \; \cos A  +  m
\;\sin A   )\; (  {d \over d r}   + {\nu \over r} \;\cos A  -  m
\; \sin A   )
   \right.
$$
$$
\left.  +   \left ( {d \over dr } \ln  {r \over \nu \sin A + e +
2m\cos A \; r } \right ) (  {d \over d r}   + {\nu \over r} \;\cos
A  -  m  \; \sin A   )   \right.
$$
$$
\left. + { e + \nu \sin A  + 2m\cos A \; r \over r}
 \; {e - \nu \sin A   \over  r}  \right ]   F   = 0 \; .
$$

\noindent After simple transformation, we arrive at

$$
\left [ {d^{2} \over dr^{2}}   +  \left ( {1 \over  r} - { 2m \cos
A    \over  \nu \sin A + e + 2m\cos A \; r } \right )
 {d \over d r} \right.
$$
$$
\left.     - { 2m \cos A    \over  \nu \sin A + e + 2m\cos A \; r
}
 \left  ( {\nu \over r} \;\cos A  -  m  \; \sin A  \right   )  \right.
$$
$$
\left.
  + E^{2} -  m^{2}   + {e^{2} - \nu^{2}  \over  r^{2}} + {2 e  E  - m \sin A   \over  r}  \right ] F = 0 \; .
$$

\noindent With special notation for the additional singular point

$$
R = - {e+ \nu \sin A \over  2m \cos A} \; ,
$$

\noindent  it reads shorter

$$
\left [ {d^{2} \over dr^{2}}   +  \left ( {1 \over  r} - { 1 \over
r - R } \right )
 {d \over d r}  + {m \sin A \over r -   R} - {\nu \cos A \over R} ({1 \over r - R} -{1 \over r})  \right.
$$
$$
\left.
  + E^{2} -  m^{2}   + {e^{2} - \nu^{2}  \over  r^{2}} + {2 e  E - m \sin A     \over  r}  \right ] F = 0 \; .
\eqno(14b)
$$

In the variable $x=r/R$,  it  looks simpler

$$
{d^{2}F\over dx^{2}}+\left[{1\over x}-{1\over
x-1}\right]\,{dF\over dx}
+\left[(E^{2}-m^{2})\,R^{2}-{\nu^{2}-e^{2}\over x^{2}} \right.
$$
$$
\left. +{-\nu \cos A+m R\, \sin A\over x-1}+{R(2eE-m \sin A)+\nu
\cos A\over x}\right]F=0\,. \eqno(14c)
$$

\noindent Let $ F=x^{a}\,e^{bx}\,\varphi(x)$, the function
$\varphi$ satisfies

$$
{d^{2} \varphi \over dx^{2}}+\left[{2a+1\over x}+2b-{1\over
x-1}\right]\,{d \varphi\over dx}
$$
$$
+ \left[b^{2}+(E^{2}-m^{2})\,R^{2}+ {a^{2}-\nu^{2}+e^{2}\over
x^{2}}-{a+b+\nu \cos A-m R\, \sin A\over x-1} \right.
$$
$$
\left. + {a+b+2ab+R(2eE-m \sin A)+\nu \cos A\over
x}\right]\,\varphi=0\,.
$$

\noindent When  $a$ and $b$ are

$$
a=\underline{+\sqrt{\nu^{2}-e^{2}}}\,, \;\; -
\sqrt{\nu^{2}-e^{2}}\; ,
$$
$$
b= +\sqrt{m^{2}-E^{2}}\,R\, , \;\;  \underline{
-\sqrt{m^{2}-E^{2}}\,R} \; , \eqno(15a)
$$

\noindent it reads

$$
{d^{2} \varphi \over dx^{2}}+\left (2b+{2a+1\over x}-{1\over
x-1}\right )\,{d \varphi\over dx}
$$
$$
+\left ({a+b+2ab+R(2eE-m \sin A)+\nu \cos A\over x} \right.
$$
$$
\left. - {a+b+\nu \cos A-m R\, \sin A\over x-1}\right )
\varphi=0\, , \eqno(15b)
$$

\noindent which is a confluent Heun equation for
 $G(\alpha,\beta,\gamma,\delta,\eta,z)$

$$
G''+\left (\alpha+{1+\beta\over z}+{1+\gamma\over z-1}\right)G'
$$
$$
+ \left ({1\over 2}\,{\alpha+\alpha \beta-\beta-\beta
\gamma-\gamma-2\eta\over z}+{1\over 2}\, {\alpha+\alpha
\gamma+\beta+\beta \gamma+\gamma+ 2\delta+2\eta\over z-1}\right )
G=0\, ,
$$
$$
\eqno(15c)
$$

\noindent with parameters

$$
\alpha=2b\,, \qquad \beta=2a\,, \qquad \gamma=-2\,,
$$
$$
\delta=2\,eER\,,  \qquad \eta=1+m R\,\sin A-2\,eER-\nu \cos A\,.
\eqno(15d)
$$

Imposing the known condition for polynomial solutions

$$
\delta=-\left (n+{\beta+\gamma+2\over 2}\right )\,\alpha\,,\qquad
n=0,1,2,... \eqno(16a)
$$

\noindent we derive  the energy quantization rule

$$
a= + \sqrt{\nu^{2}-e^{2}}\,,\qquad b= - \sqrt{m^{2}-E^{2}}\,R\,,
$$
$$
eER=(n+\sqrt{\nu^{2}-e^{2}})\,\sqrt{m^{2}-E^{2}}\; R
\qquad\Rightarrow
$$
$$
E = { m \over  \sqrt{1 +  e^{2}  /  (n + \sqrt{\nu^{2} - e^{2}}
)^{2}       }} \; , \eqno(16b)
$$

\noindent which coincides with the known result.

It should be emphasized that  confluent Heun equations in  cases
1) and  2) formally coincide,
 however   all  parameters are  different in fact:

$$
1) \qquad \alpha=2b\,, \qquad \beta=2a\,, \qquad \gamma=-2\,,
$$
$$
\delta=2\,eER\,,  \qquad \eta=1+m R\,\sin A-2\,eER-\nu \cos A\,,
\eqno(17a)
$$

$$
2) \qquad \alpha=2b\,, \qquad \beta=2a\,, \qquad \gamma=-2\,,
$$
$$
\delta=2\,eER\,,  \qquad \eta=1+m R\,\sin A-2\,eER-\nu \cos A\,,
\eqno(17b)
$$

\noindent where

$$
R = -{2e \over E + m \cos A }    \; , \qquad \sin A = {e \over
\nu} \; , \qquad \cos A = \sqrt{1 - {e^{2} \over  \nu^{2}}  } \;;
\eqno(18a)
$$
$$
R=  - {e+ \nu \sin A \over  2E }\;, \qquad \cos A = {E \over m} \;
, \qquad  \sin A = \sqrt{1 -  {E^{2} \over m^{2}} }\; .
\eqno(18b)
$$

Let  us  turn again to radial equations in presence of Coulomb
potential

$$
 ({d \over dr} \;+\; {\nu \over
r}\;) \; f \; + \; ( E  + {e \over  r} \;+ \;
  m )\; g \; = \;0 \; ,
 $$
 $$
 ({d \over dr} \; - \;{\nu \over r}\;)\; g  \;- \; ( E + {e \over r}
\; - \;
  m )\; f\; =\; 0     \; .
\eqno(19)
$$

\noindent Excluding the function  $g$, one gets

$$
{d^{2}f\over dr^{2}}+{ e \over r(E r+e +mr)}\,{df\over
dr}+\left[{e(e^{2}-\nu^{2})\over r^{2}(E r+ e+mr)}\right.
$$
$$
\left.+{ E\,(3 e^{2}-\nu^{2})-\nu\, (m+ E)+m\,(e^{2}-\nu^{2})\over
r\,(E r+e+mr)} \right.
$$
$$
\left. + {e\,(E+m)\,(3E-m)\over E r+ e +mr}+{r(E-m)(E+m)^{2}\over
E r+ e  +mr}\right]f=0\,. \eqno(20)
$$

\noindent After changing the variable
$$
x=-{(E+m)\,r\over e }\,,
\eqno(21)
$$

\noindent eq.  (20) takes the form
$$
x\,{d^{2}f\over dx^{2}}-{1\over x-1}\,{df\over dx}+\left[{ e^{2}(E
x- mx-2 E)\over E+m}+{e ^{2}-\nu^{2}\over x}-{\nu\over
x-1}\right]f=0\,. \eqno(22)
$$

\noindent Separating two factors
$$
f (x) = x^{A} e^{Cx} F (x) \; ,
$$

\noindent  for $F$ one derives

$$
{d^{2}F\over dx^{2}}+\left(2C+{2A+1\over x}-{1\over
x-1}\right){dF\over dx}+\left[C^{2}+{e^{2}(E-m)\over E+m}\right.
$$
$$
\left.+{A^{2}+ e^{2}-\nu^{2}\over x^{2}}+{A+C+2AC-2E
e^{2}/(E+m)+\nu\over x}-{A+C+\nu\over x-1}\right]F=0\,.
$$
$$
\eqno(23)
$$

\noindent When  $A, C$ are taken as  (bound states are of the
prime  interest)

$$
C^{2}+{e^{2}(E-m)\over E+m}=0\qquad\Rightarrow \qquad C= + e
\sqrt{m-E\over m+E}\,,
$$
$$
A^{2}+ e^{2}-\nu^{2}=0\qquad\Rightarrow \qquad A= + \sqrt{\nu^{2}-
e^{2}}\,, \eqno(24)
$$

\noindent eq. (23)  becomes simpler

$$
{d^{2}F\over dx^{2}}+\left(2C+{2A+1\over x}-{1\over
x-1}\right){dF\over dx}
$$
$$
+ \left[{A+C + \nu +2AC-2E e^{2}/(E +m) \over x}-{A+C+\nu\over
x-1}\right]F=0\,, \eqno(25)
$$

\noindent which is  the confluent Heun equation for  $F (\alpha,
\beta, \gamma, \delta, \eta; x)$

$$
{d^{2} \over d z^{2}} F + \left (  a + {\beta +1\over z} +
{\gamma+1 \over z-1}  \right ) {d F \over d z}
$$
$$+\left ({1\over 2}\,{a+a\gamma+\beta+\beta \gamma+\gamma+2\delta+2\eta\over z-1}
+{1\over 2}\,{a\beta+a-\beta \gamma-\beta-\gamma-2\eta\over
z}\right ) F=0
$$
$$
\eqno(26a)
$$

\noindent with parameters determined by

$$
a=2C = + 2 e \; \sqrt{m-E\over m+E} \,, \qquad \beta=2A = +2 \;
\sqrt{\nu^{2}-e^{2}} \,, \qquad
$$
$$
\gamma=- 2\,, \qquad \delta= -{2E e^{2}\over E+m}\,, \qquad
\eta=1-\nu+{2E e^{2}\over E+m}\,. \eqno(26b)
$$

The  known condition to reach polynomials   is

$$
\delta=-a\,\left (n+{\gamma+\beta+2\over 2}\right )\,.
\eqno(27a)
$$

\noindent  It results in

$$
-{2E e^{2}\over E+m} =-  2 e \; \sqrt{m-E\over m+E}  \; (  \;  n+
\sqrt{\nu^{2}-e^{2}} \;  )\,, \eqno(27b)
$$

\noindent or
$$
{E e \over  \sqrt{m^{2} - E^{2}} } = N \;, \qquad   N = n +
\sqrt{\nu^{2} - e^{2}}  \; ; \eqno(27c)
$$

\noindent from whence it follows

$$
  E  = {m  \over \sqrt{1 +e^{2} / N^{2}}} \; .
\eqno(28)
$$

\noindent It is the  exact energy spectrum for hydrogen atom in
the Dirac theory.

\section*{Supplement A. Coulomb problem in spaces of constant curvature
}

In the Lobachevsky   space $H_{3} $, the problem  of Dirac
particle in Coulomb field reduces to a  radial system (let us
specify the  case $\delta = 1$)

$$
 ( {d \over d  \beta }  +  {\nu \over
\mbox{sinh} \; \beta } \;) \; f \; + \; ( E  + { e \over
\mbox{tanh}\;\beta } \;+ \;
  m )\; g \; = \;0 \; ,
 $$
 $$
 ({d \over d \beta } \; - \; {\nu \over \mbox{sinh}\;\beta } \;)\; g  \;- \; ( E + {e \over \mbox{tanh}\;\beta }
\; - \;
   m )\; f\; =\; 0     \; .
\eqno(A.1a)
$$

\noindent From this it follows  a second order equation for $f$

$$
{d^{2} f  \over d \beta^{2} }     + { e \over \sinh \beta} { 1
\over [
 e  \cosh \beta + (E  + m ) \sinh \beta ]
}  \; {d f  \over d \chi }
$$
$$
 + \left [ (E + { e  \over
 \mbox{tanh} \;  \beta }  )^{2}   - m^{2}     - { \nu^{2} + \nu
\cosh \beta \over \sinh^{2} \beta }     + { \nu  \over \sinh^{2}
\beta }  \;\; { e  \over   e  \cosh \beta + (E  + m  )  \sinh
\beta  }
 \right ]  f = 0  \; .
$$
$$
\eqno(A.1b)
$$

Let us apply  a linear transformation with  $a (\beta) b (\beta) -
c (\beta) d(\beta)  = 1$:

$$
f (\beta) = a  \;  F (\beta) + c  \; G (\beta)\; , \qquad
 g (\beta) =  d  \; F(\beta) + b  \; G(\beta) \; ,
\; \;
$$
$$
F(\beta) = b  \; f(\beta)  - c \;  g (\beta) \; ,\qquad
  G(\beta) = - d  \; f(\beta) + a  \; g(\beta) \; .
\eqno(A.2)
$$

\noindent After combining eqs. $(A.1a)$ in the same way as in
the flat space model we obtain

$$
\left [  {d \over d \beta }  -b'a +c'd  + {\nu  (ba +cd)\over
\mbox{sinh}\; \beta }   +
 (E+{e \over \mbox{tanh} \beta} + m)
bd + (E + {e \over \mbox{tanh}\; \beta} -m)  ca  \right ]  F
$$
$$
=  \left [  b'c - b c' - {\nu \over  \mbox{sinh}\; \beta } 2bc -
(E+{e \over \mbox{tanh} \; \beta} +m) b^{2} - (E + {e \over
\mbox{tanh}\; \beta} -m) c^{2}  \right  ]  G \; ,
$$
$$
\left [  {d \over d \beta }   + d'c - a'b  - {\nu (dc  +ab) \over
\mbox{sinh}\; \beta }   - (E+{e \over \mbox{tanh}\; \beta} + m) \;
bd - (E + {e \over \mbox{tanh}\; \beta} -m)  ca  \right ]  G
$$
$$
= \left  [  -d'a +d a' + {\nu \over  \mbox{sinh}\; \beta } 2ad  +
 (E+{e \over \mbox{tanh}\; \beta} +m) d^{2} + (E + {e \over \mbox{tanh}\; \beta} -m) a^{2} \right  ]  F \; .
$$
$$
\eqno(A.3)
$$

\noindent When the above transformation is orthogonal

$$
S = \left | \begin{array}{cc}
a & c \\
d & b
\end{array} \right | =
\left | \begin{array}{cc}
\cos A/2 &  \sin A/2  \\
- \sin A/2 & \cos A/2
\end{array} \right |\,,
\eqno(A.4)
$$

\noindent eqs. (A.3)  become simpler

$$
\left (  {d \over d \beta }    + {\nu  \cos A \over \mbox{sinh}\;
\beta }   -  m  \; \sin A   \right )  F = \left ( -{A'\over 2}   -
{e \; \mbox{cosh}\; \beta  + \nu  \sin A  \over  \mbox{sinh}\;
\beta }    - E
 - m \cos A \right   )  G \; ,
$$
$$
\left (  {d \over d \beta }    - {\nu \cos A \over \mbox{sinh}\;
\beta } \;  +  m  \; \sin A   \right )  G = \left  (  +{A' \over
2}      + {e \; \mbox{cosh}\; \beta -  \nu  \sin A  \over
\mbox{sinh}\; \beta} + E  - m \cos A \right  )  F \; .
$$
$$
\eqno(A.5)
$$

Let ut translate  equations $(A.1a)$ to  a new variable

$$
\mbox{tanh} \; {\beta \over  2}   = z \; ;
$$

\noindent eqs. $(A.1a)$ will take the form
$$
{d\over dz} f +{ \nu \over z} f  - \left(  { 2 (E + m) \over
z^{2}-1} + {e(z^{2}+1)\over z(z^{2}-1)} \right)g=0\,,
$$
$$
{d \over dz} g -{ \nu  \over z} g + \left( {2(E -m) \over z^{2}-1}
+ {e(z^{2}+1)\over z(z^{2}-1)}  \right)f=0\,,
$$

\noindent or differently

$$
{d\over dz} f +{ \nu \over z} f + \left( {e  \over z} + {-E-e-m
\over z-1} + {E  - e +m \over z+1}   \right)g=0\,, \qquad \;\;
$$
$$
{d \over dz} g -{ \nu  \over z} g - \left({e  \over z} + {-E-e+m
\over z-1}+ {E  - e -m \over z+1}   \right) f=0\,. \eqno(A.6)
$$

From (A.6)  it follows a second order differential equation for
$f$

$$
{d^{2}f\over dz^{2}}+\left[{1\over z}+{1\over z-1}+{1\over z+1}-
2\,{ez+E+m\over ez^{2}+2(E+m)\,z+e }\right]\,{df\over dz}\,
$$
$$
+\left[2\,{2Ee^{2}-(E+m)\,\nu\over ez}+{-(E+e)^{2}+m^{2}+\nu\over
z-1}+ {(E-e)^{2}-m^{2}-\nu\over z+1}+{e^{2}-\nu^{2}\over z^{2}}
\right.
$$
$$
\left. + {(E+e)^{2}-m^{2}\over (z-1)^{2}}+{(E-e)^{2}-m^{2}\over
(z+1)^{2}}+ {2\,\nu\,[ez(E+m)+2\,(E+m)^{2}-e^{2}]\over
e\,[\,ez^{2}+2(E+m)\,z+e\,]}\right]f=0 \; .
$$
$$
\eqno(A.7a)
$$

\noindent It is convenient to use shortening notation
$$
{E+m \over e}= \sigma \;.
$$

\noindent Eq.  $(A.7a)$ has 6 singular points

$$
0\;, \qquad \infty\; , \qquad \pm 1,
$$
$$
 z_{1,\; 2}=  - \sigma  \pm \sqrt{ \sigma^{2}  -1 } \,, \qquad ( z_{1}z_{2} = 1 \;, \;\; z_{1}+z_{2} = -2\sigma )\;,
$$

\noindent and it reads

$$
{d^{2} f \over dz^{2}} + \left[ {1\over z}  +{1\over z-1} +
{1\over z+1} - { 1  \over  z - z_{1} } - {1 \over  z-z_{2}  }
\right]\,{df\over dz}\,
$$
$$
+\left[ {4Ee -2\sigma \,\nu\over z} - {(E+e)^{2} - m^{2}-\nu\over
z-1}+{(E-e)^{2}-m^{2}- \nu\over z+1} + {e^{2}-\nu^{2}\over z^{2}}
\right.
$$
$$
\left. + {(E+e)^{2}-m^{2}\over (z-1)^{2}}+ {(E-e)^{2}-m^{2}\over
(z+1)^{2}}+
 {A \over z - z_{1} } +
  {B \over z - z_{2} }  \right]f=0 \; ,
$$
$$
\eqno(A.7b)
$$

\noindent where

$$
2\nu {\sigma \; z  + 2 \; \sigma^{2} -1  \over (z-z_{1} ) \; ( z-
z_{2} )  }  =
  {A \over z - z_{1} } +
  {B \over z - z_{2} } \;,
  $$
  $$
A=   2\nu { \sigma z_{1} + 2 \sigma^{2} - 1  \over z_{1} -z_{2} }
\;, \qquad B = 2 \nu
 { \sigma z_{2} + 2 \sigma^{2}- 1 \over
z_{2} -z_{1} } \; . \eqno(A.8a)
$$

\noindent Let us introduce
  notation

$$
C = (E+e)^{2} - m^{2} \;, \qquad D = (E-e)^{2} - m^{2}\;,  \qquad
4Ee = C-D \; , \eqno(A.8b)
$$

\noindent then eq. $ (A.7b)$  can be presented shorter

$$
{d^{2} f \over dz^{2}} + \left ( {1\over z}  +{1\over z-1} +
{1\over z+1} - { 1  \over  z - z_{1} } - {1 \over  z-z_{2}  }
\right )\,{df\over dz}\,
$$
$$
+\left ( {C- D  - 2\sigma \,\nu\over z} - {C-\nu\over z-1}+{D -
\nu\over z+1} + {e^{2}-\nu^{2}\over z^{2}} \right.
$$
$$
\left. + {C  \over (z-1)^{2}}+ {D\over (z+1)^{2}}+
 {A \over z - z_{1} } +
  {B \over z - z_{2} }  \right )f=0 \; .
\eqno(A.9)
$$

Behavior of its solutions near 3 singular points $0, +1,-1$ is
given by

$$
x = +1 \; , \qquad f'' +{1 \over x-1} f' + C\; f = 0 \;, \qquad f
\sim z ^{\alpha}\;, \; \alpha  =  \pm \sqrt{-C}\; ;
$$
$$
x = -1 \; , \qquad f'' +{1 \over x+1} f' + D\; f = 0 \;, \qquad f
\sim z ^{\beta}\;, \; \beta  =  \pm \sqrt{-D}\; ;
$$
$$
x = 0 \; , \qquad f'' +{1 \over x} f' +(e^{2}-\nu^{2})f = 0 \;,
\qquad f \sim z ^{M}\;, \; M  = \pm \sqrt{\nu^{2} - e^{2}}\; .
$$
$$
\eqno(A.10a)
$$

Besides, near the singular point $z = \infty$ we have

$$
x = \infty  \; , \qquad f'' +{1 \over x} f' +({  A + B -2\sigma
\nu\over x}  + {C +D  + e^{2} - \nu^{2} \over x^{2}} ) \; f = 0
\;, \qquad
$$
remembering  that  $A + B -2\sigma \nu = 0 $ we get
$$
a \sim x^{N}\; , \qquad N = \pm \sqrt{C+D +e^{2} - \nu^{2}} \; .
\eqno(A.10b)
$$

Let us search solutions in the form

$$
f = x^{M} (z-1)^{\alpha} (z+1)^{\beta} \varphi \; ;
$$
eq. (A.9)  leads ro
$$
{d^{2}\varphi\over dz^{2}}+\left[{2M+1\over z}+ {2 \alpha+1\over
z-1}+{2 \beta+1\over z+1}-{1\over z-z_{1}}-{1\over
z-z_{2}}\right]\,{d\varphi\over dz}
$$
$$
+\left[{M^{2}+e^{2}-\nu^{2}\over z^{2}}+{\alpha^{2}+C\over
(z-1)^{2}}+{\beta^{2}+D\over (z+1)^{2}} \right.
$$
$$
 +
{C-D-(\alpha-\beta)\,(2M+1)-2\sigma(\nu+M)\over z}
$$
$$
 +{M+\alpha/2+\beta/2-C+ \nu+2M \alpha+ \alpha \beta\over z-1}
$$
$$
 -
{M+\alpha/2+\beta/2-D+\nu+2M \beta+ \alpha \beta\over z+1}
$$
$$
\left.+{1\over z-z_{1}}\,\left(A-{\alpha\over z_{1}-1}-{\beta\over
z_{1}+1}-{M\over z_{1}}\right)+ {1\over
z-z_{2}}\,\left(B-{\alpha\over z_{2}-1}-{\beta\over
z_{2}+1}-{M\over z_{2}}\right)\right]\, \varphi=0 \; .
$$
$$
\eqno(A.11a)
$$

Requiring

$$
M^{2} = \pm \sqrt{\nu^{2} - e^{2}}  \; , \qquad \alpha= \pm
\sqrt{-C} \; ,\qquad \beta = \pm \sqrt{-D} \; ,
$$

\noindent we arrive at a differential equation with six singular
points $(0, \infty, +1, -1, z_{1},z_{2})$

$$
{d^{2}\varphi\over dz^{2}}+\left[{2M+1\over z}+ {2 \alpha+1\over
z-1}+{2 \beta+1\over z+1}-{1\over z-z_{1}}-{1\over
z-z_{2}}\right]\,{d\varphi\over dz}
$$
$$
+\left[ {C-D-(\alpha-\beta)\,(2M+1)-2\sigma(\nu+M)\over z}\right.
$$
$$
+{M+\alpha/2+\beta/2-C+ \nu+2M \alpha+ \alpha \beta\over z-1}
$$
$$
- {M+\alpha/2+\beta/2-D+\nu+2M \beta+ \alpha \beta\over z+1}
$$
$$
\left.+{1\over z-z_{1}}\left(A-{\alpha\over z_{1}-1}-{\beta\over
z_{1}+1}-{M\over z_{1}}\right) \right.
$$
$$
 \left. +
{1\over z-z_{2}}\left(B-{\alpha\over z_{2}-1}-{\beta\over
z_{2}+1}-{M\over z_{2}}\right)\right] \varphi=0 \; .
$$
$$
\eqno(A.11b)
$$

Let us again try to perform a linear transformation of the type
(A.2) -- which results in

$$
\left [  {d \over d z }  -b'a +c'd  + {\nu \over z }  \; (ba +cd)
+
 \left( {e  \over z} + {-E-e-m \over z-1}
+ {E  - e +m \over z+1}   \right) \; \right. bd
$$
$$
\left. +  \left({e  \over z} + {-E-e+m \over z-1}+ {E  - e -m
\over z+1}   \right) \;  ca \ \right  ]  F
$$
$$
= \left [  b'c - b c' - {\nu \over z } \;  2bc - \left( {e  \over
z} + {-E-e-m \over z-1} + {E  - e +m \over z+1}   \right)  b^{2}
\right.
$$
$$
\left. - \left({e  \over z} + {-E-e+m \over z-1}+ {E  - e -m \over
z+1}   \right)   c^{2} \; \right ]  G \; ,
$$
$$
 \left [  {d \over d z }   + d'c - a'b  - {\nu \over z }\;  (dc  +ab) -
\left( {e  \over z} + {-E-e-m \over z-1} + {E  - e +m \over z+1}
\right)  \right. bd
$$
$$
\left. - \left({e  \over z} + {-E-e+m \over z-1}+ {E  - e -m \over
z+1}   \right) \;  ca  \right ]  G
$$
$$
= \left [  -d'a +d a' + {\nu \over  z}  \; 2ad  +
 \left( {e  \over z} + {-E-e-m \over z-1}
+ {E  - e +m \over z+1}   \right) d^{2} \right.
$$
$$
\left. + \left({e  \over z} + {-E-e+m \over z-1}+ {E  - e -m \over
z+1}   \right)\;  a^{2}  \right  ]  F \; . \eqno(A.12)
$$

When the transformation is orthogonal, eqs. (A.12) become
simpler

$$
\left [  {d \over d z }   + {\nu \over z }  \; \cos A  -
 \left( {e  \over z} + {-E-e-m \over z-1}
+ {E  - e +m \over z+1}   \right) \; {\sin A \over 2}   \right.
$$
$$
\left. +  \left({e  \over z} + {-E-e+m \over z-1}+ {E  - e -m
\over z+1}   \right) \;  {\sin A \over 2}  \right ]
 F
$$
$$
= \left [  -{A' \over 2} - {\nu \over z } \;  \sin A - \left( {e
\over z} + {-E-e-m \over z-1} + {E  - e +m \over z+1}   \right)
\cos^{2} {A \over 2}   \right.
$$
$$
\left. - \left({e  \over z} + {-E-e+m \over z-1}+ {E  - e -m \over
z+1}   \right) \;  \sin^{2} {A\over 2}  \right ]  G \; ,
$$
$$
\left [  {d \over d z }    - {\nu \over z }\;  \cos A  + \left( {e
\over z} + {-E-e-m \over z-1} + {E  - e +m \over z+1}   \right)
{\sin A \over 2}  \right.
$$
$$
\left. - \left({e  \over z} + {-E-e+m \over z-1}+ {E  - e -m \over
z+1}   \right) \; {\sin A \over 2} \right ]  G
$$
$$
= \left [ \; +{A'\over 2}  - {\nu \over  z}  \; \sin A   +
 \left( {e  \over z} + {-E-e-m \over z-1}
+ {E  - e +m \over z+1}   \right) \sin^{2} {A\over 2}   \right.
$$
$$
\left. + \left({e  \over z} + {-E-e+m \over z-1}+ {E  - e -m \over
z+1} \right)\;  \cos^{2} {A \over 2} \right  ] \; F \; .
$$

After simple algebraic  manipulation  we arrive at the system

$$
\left (  {d \over d z }   + {\nu  \cos A \over z }  \;   +
   {m \sin A  \over z-1}  -
 {m \sin A  \over z+1} \right)  F
$$
$$
= \left (  -{A' \over 2}  -
 {e  + \nu \sin A \over z} -  {-E-e-m  \cos A \over z-1}
- {E  - e +m \cos A  \over z+1}    \right) G \; ,
$$
$$
\left (  {d \over d z }    - {\nu  \cos A  \over z }    -
 {m  \sin A \over z-1} + {m  \sin A\over z+1}   \right) G
$$
$$
= \left (  +{A'\over 2}    +
  {e - \nu \sin A  \over z} + {-E-e+ m \cos A  \over z-1}
+ {E  - e -m \cos A \over z+1}   \right  )  F \; .
$$
$$
\eqno(A.13)
$$

\noindent In the limit of flat space model, $z << 1$, they lead to
the known formulas  (1.5).

Let us try the  variant (see Section {\bf 1.10})

$$
\sin A = e / \nu \; ; \eqno(A.14a)
$$

\noindent in this case eqs.  (A.13) read

$$
\left (  {d \over d z }   + {\nu   \cos A \over z }     +
   {m \sin A \over z-1}  -
 {m  \sin A \over z+1}     \right )  F
$$
$$
= \left ( \;  - {2\nu \over z }   \sin A - {-E-e-m  \cos A \over
z-1} - {E  - e +m \cos A  \over z+1}     \right )  G \; ,
$$
$$
\left ( \; {d \over d z }    - {\nu \cos A  \over z }  -
 {m \sin A \over z-1} + {m \sin A  \over z+1}
 \right )  G =
{(2m \cos A - 2 E) - 2e z  \over (z-1)(z+1)}
 \; F \; .
$$
$$
\eqno(A.14b)
$$

Excluding the variable $F$, we get

$$
\left (  {d \over d z }   + {\nu \cos A  \over z }     +
   {m  \sin A \over z-1}  -
 {m \sin A \over z+1}     \right )    { (z-1)(z+1) \over (2m \cos A - 2 E) - 2e z }
 $$
 $$
 \times
 \left (  {d \over d z }    - {\nu \cos A  \over z }  -
 {m  \sin A \over z-1} + {m  \sin A \over z+1}   \right) G \qquad
 $$
$$
= \left (   - {2\nu \over z }   \sin A - {-E-e-m  \cos A \over
z-1} - {E  - e +m \cos A  \over z+1}     \right ) G \; ,
$$
$$
\eqno(A.14c)
$$

\noindent from whence it follows

$$
{d^{2} \over  d z^{2}}G   +  \left ( {d \over d z } \ln {
(z-1)(z+1) \over (2m \cos A - 2 E) - 2e z } \right ) \left (  {d
\over d z }    - {\nu \cos A \over z }  -   {m\sin A  \over z-1} +
{m \sin A \over z+1}   \right)
  G
$$
$$
- \left (
 {\nu \over z }   \cos A  +
    {m \sin A \over z-1}  -
 {m \sin A \over z+1}   \right  )^{2} G -
\left (  {\nu \cos A \over z }  +
   {m \sin A  \over z-1}  -
 {m \sin A \over z+1} \right )  ' G
$$
$$
+ {2m \cos A - 2 E - 2e z  \over (z-1)(z+1)} \left (    {2\nu
\over z }   \sin A  + {-E-e-m  \cos A \over z-1} + {E  - e +m \cos
A  \over z+1}     \right )  G = 0  .
$$

\noindent It is convenient to introduce a special designation for
the singular point $z_{0} =(m \cos A -  E)/e$, then

$$
{d^{2} \over  d z^{2}}G   +
 \left (
{1 \over z-1 } + {1 \over z+1}  - { 1  \over z  - z_{0}   } \right
) {d \over d z }
$$
$$
+
 \left (
{1 \over z-1 } + {1 \over z+1}  - { 1  \over z  - z_{0}  } \right
) \left ( \;     - {\nu \cos A \over z }  -   {m\sin A  \over z-1}
+ {m \sin A \over z+1}   \right)
  G
$$
$$
- \left (
 {\nu \over z }  \; \cos A  +
    {m \sin A \over z-1}  -
 {m \sin A \over z+1}   \right  )^{2} G +
\left (  {\nu \cos A \over z^{2} }  +
   {m \sin A  \over (z-1)^{2}}  -
 {m \sin A \over (z+1)^{2}} \right )   G
$$
$$
-e\; \left  ( {1-z_{0} \over z-1} + {1 +z_{0} \over z+1} \right )
\left ( \;   {2\nu \over z } \;  \sin A  + {-E-e-m  \cos A \over
z-1} + {E  - e +m \cos A  \over z+1}     \right )  G = 0 \; .
$$
$$
\eqno(A.14d)
$$

With the use of notation

$$
m \cos A = c \; , \qquad  m \sin A = s \; , \qquad \nu \cos A
\rightarrow  \nu \;, \eqno(A.15)
$$

\noindent and taking in mind identity $\nu \sin A =  e$, eq.
$(A.14d)$ is written  shorter

$$
{d^{2} \over  d z^{2}}G   +
 \left (
{1 \over z-1 } + {1 \over z+1}  - { 1  \over z  - z_{0}   } \right
) {d \over d z }
$$
$$
+
 \left [ \left (
{1 \over z-1 } + {1 \over z+1}  - { 1  \over z  - z_{0}  } \right
) \left ( \;     - {\nu  \over z }  -   { s  \over z-1} + { s
\over z+1}   \right)
  \right.
$$
$$
\left. - \left (
 {\nu \over z }   +
    { s  \over z-1}  -
 { s \over z+1}   \right  )^{2}  +
\left (  {\nu  \over z^{2} }  +
   { s  \over (z-1)^{2}}  -
 {s \over (z+1)^{2}} \right )      \right.
$$
$$
\left. -e\; \left  ( {1-z_{0} \over z-1} + {1 +z_{0} \over z+1}
\right ) \left ( \;   {2 e \over z }  + {-E-e- c \over z-1} + {E
- e + c   \over z+1}     \right ) \right ]  G = 0 \; .
$$
$$
\eqno(A.16)
$$

Equation (A.16) reduces to

$$
{d^{2} \over  d z^{2}}G   +
 \left (
{1 \over z-1 } + {1 \over z+1}  - { 1  \over z  - z_{0}   } \right
) {d \over d z }
$$
$$
+\left[ \;   {1 \over z} {-\nu+4 \nu s z_{0}-4z_{0}^{2}e^{2}\over
z_{0}}+ {1 \over  z-z_{0}} \; {z_{0}^{2}\nu  + 2 z_{0}s-\nu \over
z_{0}(1+z_{0})(z_{0}-1)}-{\nu(\nu-1)\over z^{2}}\right.
$$
$$
\left. +{1\over z+1}\;(+ \nu-s^{2}-2 \nu
s+e^{2}+2e^{2}z_{0}-ez_{0}c-ez_{0}E+{s\over 1+z_{0}}) \;  \right.
$$
$$
\left. +{1\over z-1}\;(-\nu+s^{2}-2 \nu
s-e^{2}+2e^{2}z_{0}+ez_{0}c+ez_{0}E+{s\over 1-z_{0}})  \;  \right.
$$
$$
\left. + {-s^{2}+(1-z_{0})\,(e^{2}+ec+eE)\over (z-1)^{2}}  +
{-s^{2}+(1+z_{0})\,(e^{2}-ec-eE)\over (z+1)^{2}} \right] G=0\; .
$$
$$
\eqno(A.17)
$$

Let us  specify behavior of $G(z)$ near the singular points

$$
0,  \qquad \infty, \qquad +1, \qquad   -1 \; .
$$

Near the point  $z=0$ we have

$$
{d^{2} \over  d z^{2}}G   +{1 \over z_{0}} {d \over d z }G
-{\nu(\nu-1)\over z^{2}}  G=0 \; ;
$$

\noindent therefore $G$ behaves in accordance with

$$
G \sim  z^{a} ,  \qquad a(a-1) z^{a-2} +{1 \over z_{0}}  + a \;
z^{a-2} z - \nu(\nu-1) z^{a-2} = 0\qquad \Longrightarrow
$$
$$a(a-1) = \nu(\nu-1) \;, \qquad  a =  \underline{+ \nu,} \; 1 - \nu \; ;
\eqno(A.18a)
$$

\noindent positive values  $a = + \nu$ correspond to bound states.

Near the point  $z=1$, we have

$$
{d^{2} \over  d z^{2}}G   + {1 \over z-1 } {d \over d z } +
{-s^{2}+(1-z_{0})\,(e^{2}+ec+eE)\over (z-1)^{2}} \; G=0 \; ,
$$

\noindent that is
$$
G \sim (z-1)^{M}\; , \qquad M = \pm  \sqrt{ s^{2} -
(1-z_{0})\,(e^{2}+ec+eE) } \; ; \eqno(A.18b)
$$

\noindent because  $z = \mbox{tanh}\; (\beta /2)$, the sign
\underline{minus} in $(A.18b)$ corresponds to
 the bound states.

In the same manner, consider  the vicinity of the point  $z=-1$:

$$
{d^{2} \over  d z^{2}}G   + {1 \over z+1} {d \over d z } +
 {-s^{2}+(1+z_{0})\,(e^{2}-ec-eE)\over (z+1)^{2}}\; G=0 \; ,
$$

\noindent that is
$$
G \sim (z+1)^{N}\; , \qquad N = \pm  \sqrt{ s^{2}
-(1+z_{0})\,(e^{2}-ec-eE)  } \; ; \eqno(A.18c)
$$

\noindent because  $z = \mbox{tanh}\; (\beta /2)$, the singular
point $z=-1$ is not physical one.

For brevity, let us introduce special notation for coefficients at
the for singular terms

$$
{K \over z - z_{0}}\;, \qquad {K _{0}\over z} \; , \qquad
{K_{-}\over x-1} \; , \qquad {K_{+}\over x+1}  \;,
$$

\noindent then eq.  (A.17) reads  shorter
$$
{d^{2} \over  d z^{2}}G   +
 \left (
{1 \over z-1 } + {1 \over z+1}  - { 1  \over z  - z_{0}   } \right
) {d \over d z } G
$$
$$
+\left[ \;   {K_{0} \over z} + {K \over  z-z_{0}}  -{\nu
(\nu-1)\over z^{2}}+ {K_{+}\over z+1} +  {K_{-}\over z-1}
 - {M^{2}\over (z-1)^{2}}  -
{N^{2} \over (z+1)^{2}} \right] G=0 \; .
$$
$$
\eqno(A.19)
$$

Let us introduce the following substitution
$$
G = z^{a} (z-1)^{M} (z+1)^{N} \varphi \; ;
$$

\noindent after simple calculation we arrive at

$$
{d^{2}\varphi\over dz^{2}}+\left[{2a\over z}+{2m+1\over z-1}+
{2n+1\over z+1}-{1\over z-z_{0}}\right]\,{d \varphi\over dz}
$$
$$
+\left[{a(a-1)-\nu(\nu-1)\over z^{2}}+{m^{2}-M^{2}\over (z-1)^{2}}
\right.
$$
$$
+ {n^{2}-N^{2}\over (z+1)^{2}}+{1\over z-z_{0}}\,
\left(K_{0}-{a\over z_{0}}-{m\over z_{0}-1}-{n\over
z_{0}+1}\right)
$$
$$
+{1\over z}\,\left(K+{a[-2mz_{0}+1+2nz_{0}]\over z_{0}}\right)
$$
$$
+ {1\over z-1}\,\left(K_{-}+{1\over 2}\,(2a+n)\,(2m+1)+{1\over
2}\,{m(z_{0}+1)\over z_{0}-1}\right)
$$
$$
\left.+{1\over z+1}\,\left(K_{+}- {1\over
2}\,(2a+m)\,(2n+1)-{1\over 2}\,{n(z_{0}-1)\over
z_{0}+1}\right)\right]\varphi=0\,.
$$
$$
\eqno(A.20a)
$$

Requiring

$$
a(a-1)=\nu(\nu-1) \;, \qquad m^{2} =M^{2} \;, \qquad n^{2}=N^{2}
\; ,
$$

\noindent we arrive at

$$
{d^{2}\varphi\over dz^{2}}+\left[{2a\over z}+{2m+1\over
z-1}+{2n+1\over z+1}-{1\over z-z_{0}}\right]\, {d \varphi\over dz}
$$
$$
+\left[ {1\over z-z_{0}}\,\left(K_{0}-{a\over z_{0}}-{m\over
z_{0}-1}-{n\over z_{0}+1}\right)+ {1\over z}\,\left( K+{a
(-2mz_{0}+1+2nz_{0}) \over z_{0}}\right)  \right.
$$
$$
\left.+{1\over z-1}\,\left(K_{-}+{1\over
2}\,(2a+n)\,(2m+1)+{1\over 2}\,{m(z_{0}+1)\over
z_{0}-1}\right)\right. \qquad
$$
$$
\left.+{1\over z+1}\,\left(K_{+}- {1\over
2}\,(2a+m)\,(2n+1)-{1\over 2}\,{n(z_{0}-1)\over
z_{0}+1}\right)\right]\varphi=0\,.
$$
$$
\eqno(A.20b)
$$

In the Riemann space, the Kepler problem can be treated in a
similar way.

\end{document}